\documentclass[fleqn,usenatbib]{mnras}
\usepackage{newtxtext,newtxmath}

\usepackage[T1]{fontenc}

\DeclareRobustCommand{\VAN}[3]{#2}
\let\VANthebibliography\thebibliography
\def\thebibliography{\DeclareRobustCommand{\VAN}[3]{##3}\VANthebibliography}

\usepackage{graphicx}	
\usepackage{amsmath}	
\usepackage[normalem]{ulem}
\usepackage{xcolor}

\title[The evolutionary path of void galaxies]{The evolutionary path of void galaxies in TNG300 simulation}

\author[Agust\'in. M. Rodríguez-Medrano et al.]{%
Agust\'in M. Rodríguez-Medrano$^{1,2}$,\thanks{E-mail: arodriguez@unc.edu.ar (AMRM)}
Volker Springel$^{3}$,
\newauthor{
Federico A. Stasyszyn$^{1,2}$,
Dante J. Paz$^{1,2}$}
\vspace*{0.1cm}\\%
$^{1}$Instituto de Astronom\'ia te\'orica y experimental-Conicet, Laprida 854, Córdoba, Argentina\\%
$^{2}$Observatorio Astronómico de Córdoba, Universidad Nacional de Córdoba, Laprida 854, X5000BGR Córdoba, Argentina\\%
$^{3}$Max-Planck-Institut für Astrophysik, Karl-Schwarzschild-Straße 1, 85741 Garching, Germany}

\date{Accepted XXX. Received YYY; in original form ZZZ}

\pubyear{2015}

\begin{document}
\label{firstpage}
\pagerange{\pageref{firstpage}--\pageref{lastpage}}
\maketitle

\begin{abstract} 
The properties of galaxies in low-density regions of the universe suggest an interplay between galaxy formation and environment. However, the specific reason why this particular large-scale environment influences the evolution of galaxies remains unclear. This paper examines the properties and evolutionary paths of galaxies within cosmic voids using the Illustris TNG300 simulation. The population of void galaxies at $z=0$ has a higher star formation rate, a smaller stellar-to-halo-mass ratio, higher gas metallicity, and lower stellar metallicity in comparison with non-void galaxies at fixed stellar mass. Our analysis shows that these differences are mainly due to the characteristics of galaxies classified as satellites, for which the largest differences between void and non-void samples are found. 
Although the mean number of mergers is similar between void and non-void samples at a fixed stellar mass, void galaxies tend to experience mergers at later times, resulting in a more recent accumulation of accreted stellar mass. While the mean net accreted mass is comparable for high mass galaxies, low mass void galaxies tend to 
exhibit higher fractions of accreted stars than non-void galaxies. This finding challenges the common notion that void galaxies predominantly experience growth with infrequent mergers or interactions. With this paper, we also publicly release our void catalogue as part of the IllustrisTNG project.  
\end{abstract}


\section{Introduction}

The currently leading cosmological model, known as $\Lambda \rm CDM$, has been successfully tested against several large-scale structure properties detected in galaxy surveys \citep{Colless2003,Tegmark2004}. Galaxies in the universe group together to form the cosmic web, which can be understood as a network of interconnected structures such as nodes, filaments, walls, and voids \citep{Bond1996,Cautun2014,Libeskind2018}. The cosmic web can be used to classify the large-scale structure, and plays an important role in the evolution of the universe as well as in influencing galactic properties \citep{Martizzi2020,Nandi2023}. Since \citet{Dressler1980} until the present, numerous studies have developed theories on how the environment and dynamics of large-scale structure affect the evolution of galaxies \citep{Coldwell2002,Martinez2006,Omill2008,Duplancic2018,Rost2020,Duplancic2020,Einasto2022}.

In this context, cosmic voids constitute the most under-dense regions in the universe. The identification of these regions is non-trivial because it depends on how they are defined in terms of shape, density thresholds and tracers used. There are different ways of defining them in the literature \citep[see for instance][]{Padilla2005, Platen2007,Neyrinck2008,Lavaux2010,Sutter2015}, but overall all of them yield similar characteristics for galaxies within these voids.

The population of void galaxies tends to be dominated by low-mass, blue, star-forming galaxies with young stellar populations \citep[see for instance:][]{Rojas2004,Rojas2005,Hoyle2005,Hoyle2012,Tavasoli2015,Moorman2016,Florez2021,Jian2022}.
One explanation for this is that galaxies in these environments might encounter lower gravitational forces as a result of the lower density of matter, which may lead to slower star formation and thus smaller stellar masses. In a recent study by \citet{Dominguez2023a}, the star formation histories (SFHs) of void galaxies from the \textsc{CAVITY} project were constructed and analyzed. The findings indeed reveal that void galaxies exhibit slower SFHs compared to galaxies residing in denser environments, such a clusters, filaments, and walls.

Another possible consequence of residing in void environments is that these galaxies should also be more solitary, have fewer nearby galaxies, and have less gas accessible for star formation. So, compared to galaxies in denser regions, these galaxies may have distinct morphologies and chemical compositions. Therefore, the information on the formation and evolution of galaxies, as well as the large-scale structure of the universe can be better understood by looking at the characteristics of galaxies in cosmic voids.
The different merger rates between galaxies in cosmic voids and denser environments, have been suggested to play a role in shaping the galaxy evolution \citep{Alfaro2020, RosasGuevara2022}. In this work, we aim to conduct a detailed study of the evolution of galaxies in voids over time using the IllustrisTNG 300 simulation.

In a previous study  \citep{RodriguezMedrano2022}, we examined a set of zoom-in simulation of cosmic voids and found that haloes within voids exhibit a slower rate of evolution compared to those in denser environments. Furthermore, our analysis indicated that the rate of evolution of haloes appears to depend on the specific large-scale environment in which the void is embedded \citep[relating to the
void-in-void and void-in-cloud concepts as defined by][]{Ceccarelli2013}.  Building upon these findings,  \citet{RodriguezMedrano2023} detected distinctive signals in the properties of galaxies that suggest different evolutionary paths for galaxies within the two types of voids. Moreover, our research has highlighted that galactic evolution is not solely determined by the local environment. In this work, we aim to conduct a detailed study of the evolution of galaxies in the inner regions of voids over time in order to shed more light on this. Note that these earlier works were carried out with a spherical void definition; in the present work we will extend the study of galaxies to free-form voids identified with the \textsc{Popcorn} algorithm \citep{Paz2023}.

The differences in galaxy evolution between cosmic voids and denser environments are important for several reasons. One of these is the phenomenon known as assembly bias, which is related to the dependence of the properties of halos on their environment \citep{MonteroDorta2023}. By studying the evolution of halos in cosmic voids, we can gain insights into the mechanisms that drive assembly bias and the relationship between the large-scale structure of the universe and the distribution of galaxies within it. Additionally, the study of galaxy evolution in cosmic voids can provide valuable tests of the $\Lambda \rm CDM$ paradigm, which is the current standard model of cosmology. This includes testing the predictions of dark matter simulations and the possibility of modifying gravity on large scales. Understanding the environmental factors that influence galaxy evolution is essential for developing a comprehensive understanding of the structure and evolution of the universe \citep{Cataldi2022}.

This paper is organised as follow: in Section~\ref{sec:data} we introduce the simulation and void finder utilised, in Section~\ref{sec:results} we present the results about the properties of void and non-void galaxies, first at $z=0$ and then for the evolution of the properties with redshift. In Section~\ref{sec:discussions} we discuss our results in the context of the background in the literature. Finally, in Section~\ref{sec:conclusions} we present the main conclusions of this work.

\section{Simulation and void identification}
\label{sec:data}
\subsection{Simulation}

The IllustrisTNG simulations \citep{Marinacci2018,Naiman2018,Nelson2018,Pillepich2018,Springel2018} are a set of  magnetohydrodynamical simulations performed with the AREPO code \citep{Springel2010}. In this work we used the largest volume simulated as part of the project that consists of a box with $\sim 300 $ comoving Mpc. The IllustrisTNG simulations adopt a $\Lambda$CDM cosmology \citep{PlanckCollaboration2016}. The cosmological parameters used are $\Omega_m=0.3089$, $\Omega_b=0.0486$ , $\Omega_{\Lambda}=0.6911$, $h=0.6774$ , $\sigma_8=0.8159$, and $n_s=0.9667$. The IllustrisTNG-300 simulation follows the evolution of $2500^{3}$ dark matter particles and the same quantity of gas cells, and reached a mass resolution of $\sim 6\times10^{7}{\rm M}_{\odot}$ and $1\times10^{7}{\rm M}_{\odot}$ for dark matter and gas, respectively.

The simulation uses subgrid models to represent a variety of physical processes, like star formation, radiative cooling, chemical enrichment by supernovae, seeding of magnetic fields, growth of supermassive black holes (BHs), and feedback by BHs and stars. The models used were calibrated to reproduce the galaxy stellar mass function at $z=0$, and they simultaneously provide reasonable matches of the halo gas fraction, the galaxy stellar size distribution, the cosmic SFR density, and the relation between the black hole and galaxy mass \citep{Weinberger2017,Pillepich2018}.

The haloes and subhaloes were identified with the SUBFIND algorithm \citep{Springel2001,Dolag2009}, and the merger trees we use here were constructed with SUBLINK \citep{RodriguezGomez2015}. The classifications of central or satellite galaxies are provided by SUBFIND, where the central galaxy is defined as the subhalo with the highest number of bound particles/cells within a FOF group, and the rest of the galaxies are satellites.
The code identified galaxies as dark matter subhaloes together with the gas and stellar particles they contain. In this work, we use a cut of $10^{8.5}\,{\rm M}_{\odot}$ in stellar mass for our galaxies. This cut is adopted to mitigate numerical resolution effects, as for these masses galaxies are composed of $\sim 50$ stellar particles.  

However, it is important to note that the IllustrisTNG simulations quantitatively resolve galaxy population properties well only for galaxies with at least 100 stellar particles \citep{Pillepich2018}. 
Therefore, the low-mass end of our galaxies may face certain resolution issues, in particular for galaxies with less than 100 stellar particles.  We expect, however, that these systematics are similarly present in both void and high-density environments for a fixed stellar mass, still allowing a qualitative comparison between galaxies in different environments. In the following sections, we will show that the behaviour of the various astrophysical properties analysed in this work remains qualitatively consistent when considering a minimum of 50 stellar particles per galaxy compared with stellar masses equivalent to 100, 300, or 500 particles.

\subsection{Void Identification}

Void regions are identified using the public version of the \texttt{Popcorn} void finder\footnote{The source code is available at
\url{https://gitlab.com/dante.paz/popcorn_void_finder} under the MIT license.} \citep{Paz2023}.  The algorithm is a generalisation
of the spherical void finder \citep[SVF, see][for more details]{Paz2023}, so let us introduce this method first. 
For a given tracer sample (particles, galaxies, or halos), the density field is first estimated and all locally underdense regions 
are seeded. A sphere is then grown around each seed, searching for the largest radius where the integrated density contrast is below a given threshold (in our case, $\Delta=-0.9$). After expanding all the seeds, spheres that are touching each other are cleaned by favouring the larger ones. In a hierarchical clustering scenario, the abundances of spherical voids (SV) should be described as a monotonic decreasing function of their size, exhibiting a behaviour similar to a Press-Schechter function. However, void counts deviate from this behaviour below a certain scale, which we refer to as the shot noise radius ($R_\mathrm{shot}$). This radius represents the minimal measurable size of voids in the simulation without being affected by the tracer discreteness. The full Popcorn void finder algorithm consists of a series of steps:
\begin{enumerate}
 \item First, the SVF is run, and the $R_\mathrm{shot}$ scale is measured.
 \item Later, for each SV, its surface is regularly covered with seeds (following a Fibonacci covering scheme).
 \item \label{pt3} Each seed is then expanded one at a time, searching for the largest radius of a second sphere within the joint volume, where the density contrast is below a given threshold. The largest radius of each seed is recorded, and the seed that expanded the most is identified.
 If this radius is larger than $R_\mathrm{shot}$, the sphere member is accepted; otherwise, the process ends.
\item If the sphere is accepted, it becomes part of the popcorn candidate, and the entire surface of the object is covered with seeds again. \item The procedure is repeated from step \ref{pt3}, searching for the maximal third sphere that expands the popcorn volume the most while the contrast stays below the density threshold. The iteration ends when no larger sphere than $R_\mathrm{shot}$ can be added.
\item As a final step, all overlapping popcorn candidates having an intersecting volume larger than $4\pi R_\mathrm{shot}^3/3$ are removed, favouring those with the largest starting SV.

In this way, we identify voids over the haloes with $M_{\rm tot}>3\times10^{11} \rm M_{\odot}$.
We find 474 void regions containing 15391 galaxies. In Figure~\ref{fig:voids}, we present a slice of the simulation with the void regions. The points indicate all the subhaloes in the simulation slice. We observe several voids with an abundance of points, however, we have to emphasize that we only consider haloes with a minimum mass for identifying the voids. The \texttt{Popcorn} algorithm was designed to avoid the problem of fragmentation of void regions and the associated impact on the estimate of void abundances and thus the comparison with excursion set models. In this work we adopted the \texttt{Popcorn} void finder for defining underdense regions due to its ability of a more complete covering of the void regions, allowing larger samples of galaxies inside voids than those detected using a spherical void finder. However, all the trends presented in this work have also been obtained using our spherical void finder, albeit with slightly larger errors. In this way we expect the results presented here can be compared with those using a similar spherical void finder \citep[see for instance][]{Padilla2005,Ruiz2015,Paillas2017}.

\begin{figure}
	\includegraphics[width=\columnwidth]{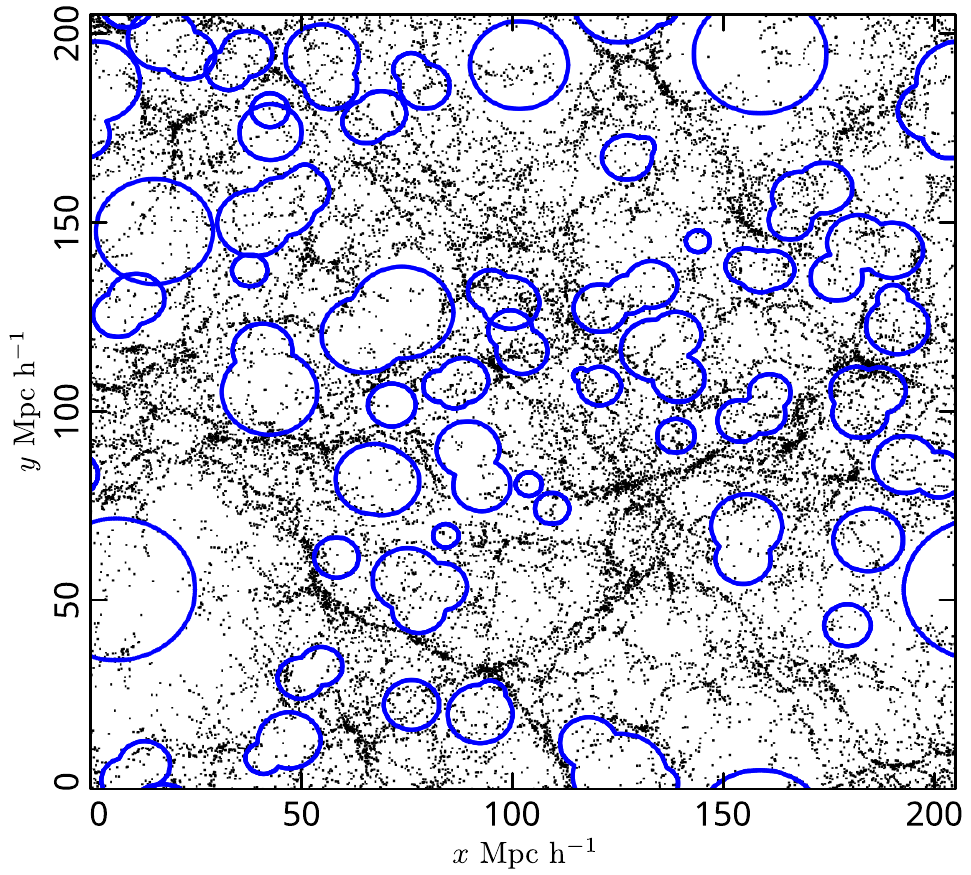}
    \caption{Slice with a depth $20 \,h^{-1}\rm Mpc$ of the TNG300 simulation. The points indicate the subhaloes in the slice, and the circles the void regions identified for haloes with $M_{\rm tot}>3\times10^{11} {\rm M}_{\odot}$.}
    \label{fig:voids}
\end{figure}

\end{enumerate}

\section{Data analysis}
\label{sec:results}

In this section, we explore the astrophysical properties of galaxies residing within cosmic voids and compare them with to non-void galaxies. First, we analyse present time galaxy properties in subsection~\ref{sec:results_z=0}, then subsequently extend this analysis to higher redshifts in subsections~\ref{sec:results_evolution} and \ref{sec:results_evolution2}. We conclude this section by carrying out an analysis of the mergers and accreted masses of our galaxies in subsection~\ref{sec:results_mergers}.

\subsection{Galaxies at the present time}
\label{sec:results_z=0}
\begin{figure}
	\includegraphics[width=\columnwidth]{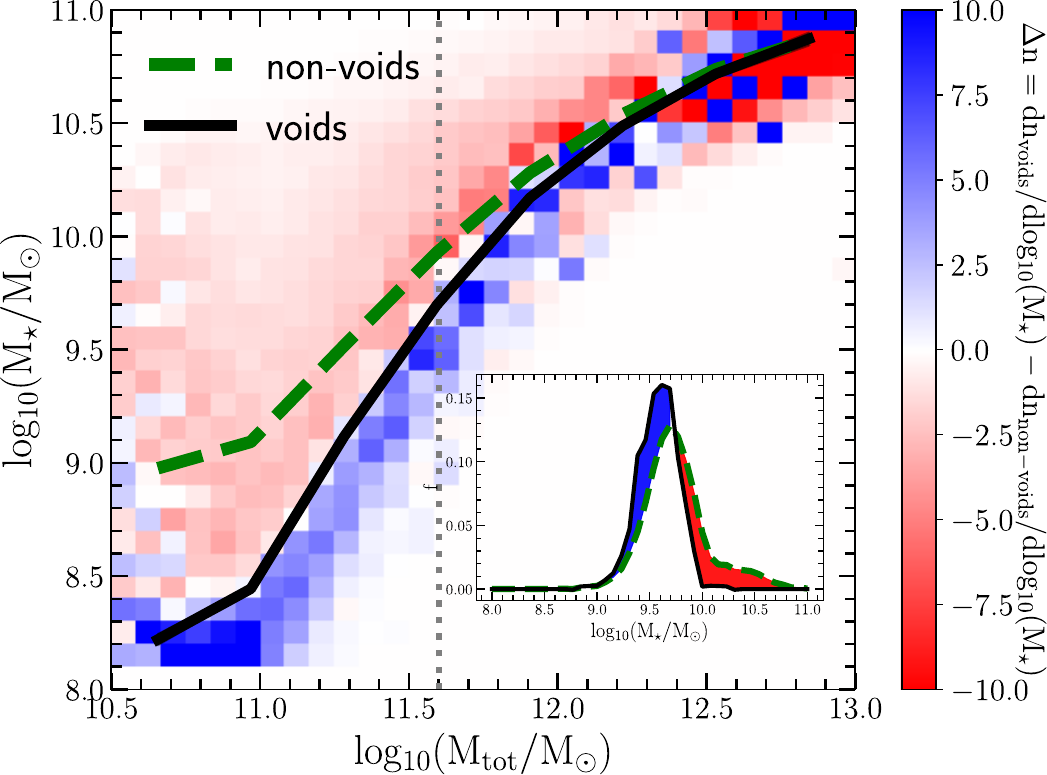}
    \caption{The lines show the mean stellar to halo mass relation for galaxies in voids (solid black line) and non-void galaxies (dashed-green line) as a function of the total mass to stellar ratio. 
    The vertical grey shaded line marks the cut of $\log(M_{\rm tot}/{\rm M}_{\odot}) = 11.6$, and the distribution of $M_{\star}$ for the galaxies is shown in the inset panel. For this cut, the distribution of $M_{\star}$ shows an excess of galaxies with $\log(M_{\star} /{\rm M}_\odot) <9.8$ for the void sample. This excess is represented in blue in the colour map, whereas the deficit is shown in red.
    The colour map shows the differences between stellar mass distributions for galaxies in void and non-void regions, at a fixed total mass (see for instance the inset panel).}
    \label{fig:abundance}
\end{figure}

As a starting point, we focus on the stellar-to-halo mass relation (SHMR). The mean relation between the stellar mass and halo mass for each subhalo in the simulation is depicted in Figure \ref{fig:abundance}. The masses are calculated by \texttt{SUBFIND} as the sum of the masses of all particles and cells gravitationally bound to each subhalo. We present the mean relation for non-void galaxies with the dashed-green line, while the solid-black line represents the relation for galaxies within voids. 
We calculated the errors for the means using the jackknife technique. These were represented as a shaded area on the mean of the void sample, which has the highest error. However, due to the low error value, it is not visible in the figure.
A comparison between the two samples reveals that galaxies within voids exhibit a smaller SHMR for galaxies with $M_{\rm tot}<10^{12.5}\,{\rm M}_{\odot}$. We compute for bins in $M_{\rm tot}$ the stellar mass distribution of galaxies inside and outside void regions. The colour map in the figure represents the arithmetic difference between both distributions, indicating with bluer colours positive values and a larger frequency of galaxies in voids, while redder colours indicate a lower frequency of galaxies in voids (negative values), i.e.~${\rm d}n_\mathrm{void}/{\rm d}\mathrm{log}(M_{\star}) - {\rm d}n_\mathrm{non-void}/{\rm d}\mathrm{log}(M_{\star})$ at a fixed total mass. To exemplify this procedure, the inset panel of the figure shows the stellar mass distribution for void galaxies and the general sample. These distributions correspond to galaxies within the halo-mass bin indicated by the vertical dashed-black line, i.e. $10^{11.25}\,{\rm M}_{\odot}$. The inset panel demonstrates that galaxies in voids exhibit an excess in the distribution for $M_{\star}<10^{9.8}\,{\rm M}_{\odot}$ and a scarcity for $M_{\star}>10^{9.8}\,{\rm M}_{\odot}$. Thus, the figure not only highlights the variation in the mean SHMR based on the galactic environment but also shows how these variations arise from trends in the stellar mass distribution when the total mass is fixed.

The dependence of the SHMR with respect to the cosmic environment has been reported in previous papers, such as \citet{Alfaro2020, Habouzit2020,RosasGuevara2022} for void galaxies, and \citet{Martizzi2020} for general large-scale environments. Numerical simulations consistently reveal that galaxies within voids exhibit a reduced stellar content when compared to galaxies in the overall universe, while maintaining a fixed halo mass. These results suggest that galaxies per stellar mass bin in voids evolve in a relatively higher gravitational potential well (i.e.~larger halo masses for objects in voids), which may impact their galactic properties. 

As depicted in Figure \ref{fig:abundance}, halos with masses of approximately $M_{\rm tot}\sim 3 \times 10^{11}$ correspond to galaxies with masses around $M_{\star}\sim 10^{9.5}$. This mass is associated with the identification of voids; thus, beyond this mass, potential biases may arise in the results. To mitigate these effects in this study, we consider galaxies with $M_{\star}<10^{10}$, where this threshold is slightly larger to account for the dispersion in the stellar-to-halo mass relation (SHMR).

\begin{figure}
	\includegraphics[width=\columnwidth]{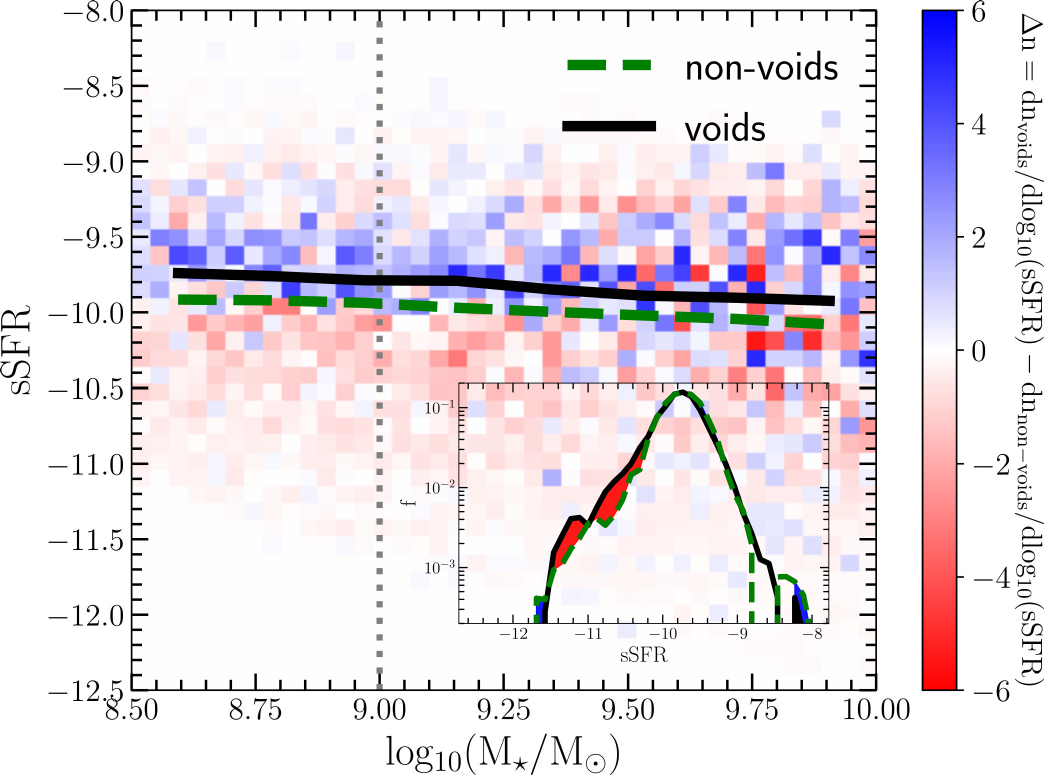}
    \caption{The lines show the mean relation between the specific star formation rate (sSFR) and stellar mass for void galaxies (green line) and all the sample (black shaded line). In order to highlight the differences in the distributions of sSFR we show (as in Fig. \ref{fig:abundance}) the excess of void-galaxies in the sSFR distribution. }
    \label{fig:ssfr_mst}
\end{figure}

The star formation rate (SFR) is often reported to be higher in void environments \citep{Rojas2005, Ricciardelli2014, Habouzit2020, RodriguezMedrano2023}. In Figure \ref{fig:ssfr_mst}, we present 
the relation between the specific star formation rate (sSFR), defined as $\rm log_{10}(SFR/M_{\star})$, and the stellar mass. We calculate this relation only for galaxies with ${\rm sSFR}>-12.5$ to avoid considering galaxies where numerically the SFR could not be calculated (i.e.~galaxies with vanishing star formation). The solid-black and dashed-green lines represent the mean relation for the void and non-void samples, respectively. 
As in Fig. \ref{fig:abundance}, the errors in the mean were calculated using the jackknife technique. The voids sample has the highest magnitude for errors and they were represented in the figure as shaded areas. However, due to their low values, they do not become visible.
The colour map in this figure indicates the excesses and scarcities observed in the sSFR distribution for void galaxies compared to the general sample (similarly to Fig. \ref{fig:abundance}, the colour map encodes distribution differences). The inset panel provides an example of the sSFR distribution at a given stellar mass cut ($10^9\,{\rm M}_{\odot}$), as indicated by the vertical line. The figure demonstrates a tendency for void galaxies to exhibit higher sSFR over the full stellar mass range. Although this trend may appear weak when considering the mean of the distribution, the colour map reveals its consistent presence across the indicated stellar mass range.


\begin{figure}
	\includegraphics[width=\linewidth]{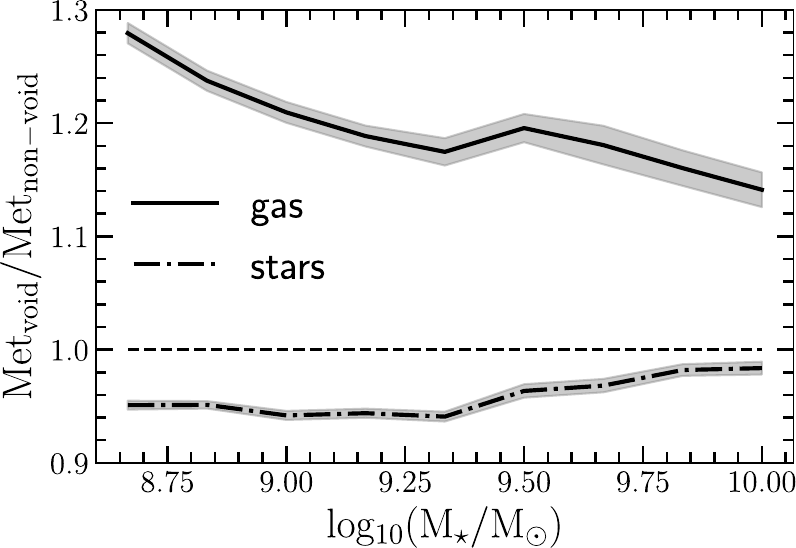}
    \caption{Average metallicity ratio between galaxies in void and non-void environments, as a function of the logarithm of stellar mass. A solid line represents the gas metallicity, while a dot-dashed line shows the stellar metallicity. All galaxies are identified at $z=0$. }
    \label{fig:metalicidades}
\end{figure}

The relatively higher gravitational potential wells in which void galaxies reside, compared to non-void galaxies, could result in a more efficient retention of metal-enriched gas expelled through feedback processes. This scenario was proposed by \citet{RosasGuevara2022} as a plausible explanation for the elevated metallicity signals detected in void galaxies. To further investigate this hypothesis, we present the metallicity ratio between void and non-void galaxies as a function of stellar mass in Figure \ref{fig:metalicidades}. A solid line represents the gas metallicity, while the metallicity of star particles is denoted by a dashed line. The shaded area indicates the error in the mean. In both cases, the calculations are confined to cells within twice the stellar half-mass radius. The Figure~\ref{fig:metalicidades} shows that, for galaxies with stellar masses $M_{\star}<10^{10.5}\,{\rm M}_{\odot}$, the gas in void galaxies exhibits higher metallicities compared to the non-void sample. On the other hand, the metallicity of stellar particles in non-void galaxies shows a slight excess in comparison to the void sample. Although the differences, in this case, are smaller in magnitude, they remain consistent within the aforementioned stellar mass range. These findings suggest that the higher potential wells associated with void galaxies may effectively retain gas enriched with metals more efficiently, offering a plausible mechanism to explain the abundance of metal-rich gas observed in these galaxies.
For a given stellar mass, galaxies in larger haloes have galaxies with smaller stellar metallicity and age \citep{Scholz2022}. This is consistent with our results and, in turn, suggests that the galaxies in the void sample would have younger ages.
Previous studies have suggested that void galaxies tend to be younger than their counterparts in denser environments \citep{Rojas2005,Ricciardelli2014,Alfaro2020,RodriguezMedrano2022}.

\subsection{Evolution of galaxies}
\label{sec:results_evolution}

Figures~\ref{fig:abundance} and \ref{fig:ssfr_mst} reveal two significant characteristics of void galaxies. The properties of galaxies within voids appear to be influenced by the galactic environment, particularly for galaxies with stellar masses of approximately $M_{\star}\lesssim 10^{10.2}\,{\rm M}_{\odot}$. To highlight how the environment contributes to these differences in galactic and halo properties, we examine the formation of haloes in both galaxy samples, as well as the growth of their galaxies. To achieve this, we analyse the merger trees derived from the simulation data. We focus on a specific stellar mass bin where the properties of galaxies exhibit an environmental dependence. However, it should be noted that galaxies in the low-mass regime have a limited number of particles at $z=0$, making it difficult to trace their evolution extensively in the merger trees. To overcome this limitation, we select all galaxies within the mass range $9.5<\log(M_{\star}/{\rm M}_\odot) <10$ at $z=0$ to ensure a robust particle sample size. Subsequently, we track the main branch of these galaxies across the merger trees. We have 1776 galaxies for the void sample at $z=0$ and 68236 for the non-void sample in this mass bin. We show the results for this specific mass bin in this section. Nevertheless, it is crucial to emphasise that our findings remain consistent across a wider range of stellar masses. In Section~\ref{appendix} we extend the results to a smaller mass bin.

\begin{figure}
	\includegraphics[width=\linewidth]{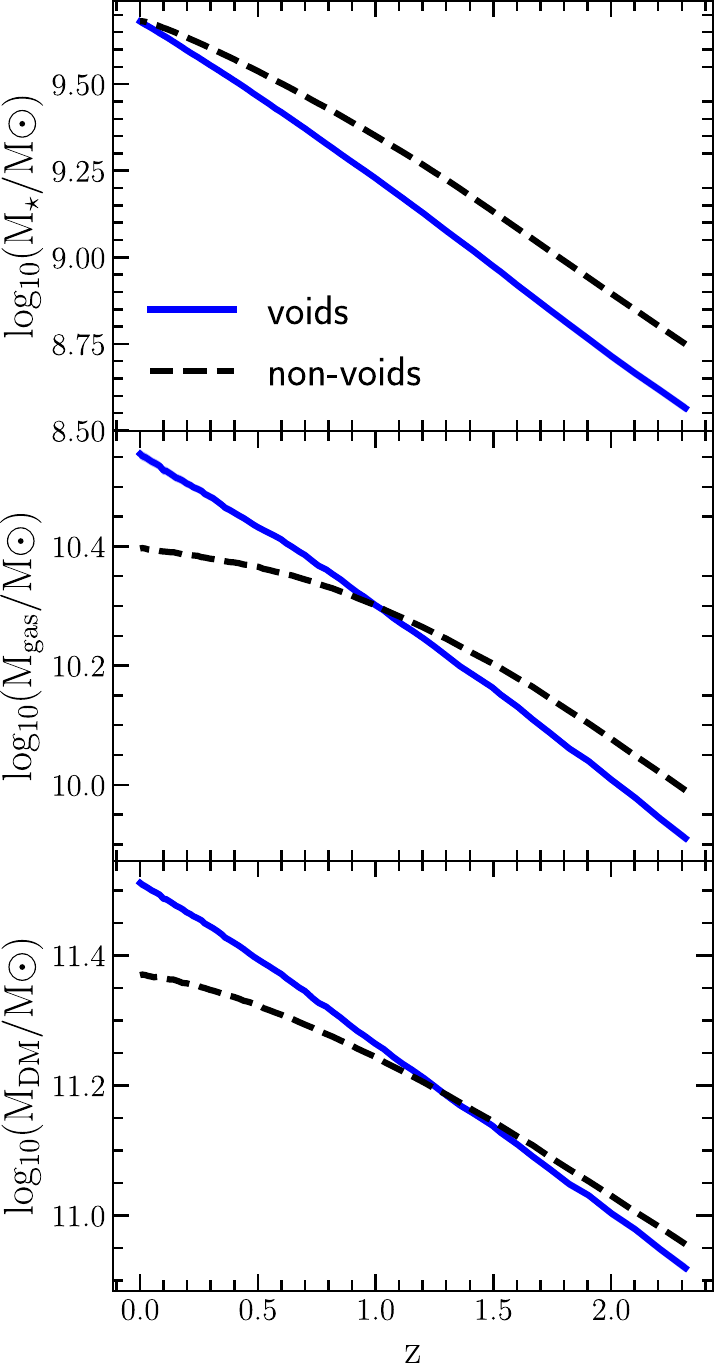}
    \caption{Evolution of different mass components with redshift for galaxies in void and non-void environments, identified at $z=0$ with $10^{9.5}\, {\rm M}_{\odot} <M_{\star} < 10^{10}\, {\rm M}_{\odot}$. \textit{Upper panel:} Average stellar mass evolution. \textit{Middle panel:} Average gas mass evolution. \textit{Bottom panel:} Average dark matter mass. In all panels, the solid blue line represents the behaviour of void galaxies, while the dashed black line is for non-void galaxies. The errors, calculated using a jackknife resampling technique, are relatively small and are depicted as shaded areas for the void sample.}
    \label{fig:masshistory}
\end{figure}

In Figure \ref{fig:masshistory}, the top panel shows the mean stellar mass evolution for galaxies with $10^{9.5}<\log(M_{\star}/{\rm M}_\odot)<10^{10}$ at $z=0$. The solid-blue line represents void galaxies, while the dashed-black line corresponds to the non-void sample. At $z=0$, both samples exhibit the same mean mass due to the narrow mass bin selection. However, as redshift increases, we observe different growth rates between the two samples. At higher redshifts, void-galaxies have, in the mean, less stellar content. In the middle panel, we present the evolution of the mean gas mass for the same galaxy population. At $z=0$, void galaxies have higher gas masses.  Similarly, the bottom panel shows the evolution of mean dark matter mass. At $z=0$, void galaxies exhibit larger dark matter masses. The excess of dark matter and gas masses explains the observed differences in the stellar-to-halo mass relation (Fig. \ref{fig:abundance}). In both cases, it is evident that void galaxies display a higher abundance of dark matter and gas content at redshifts approximately below 1. However at early times, void galaxies exhibit lower dark matter and gas content.
 
\begin{figure}
	\includegraphics[width=\linewidth]{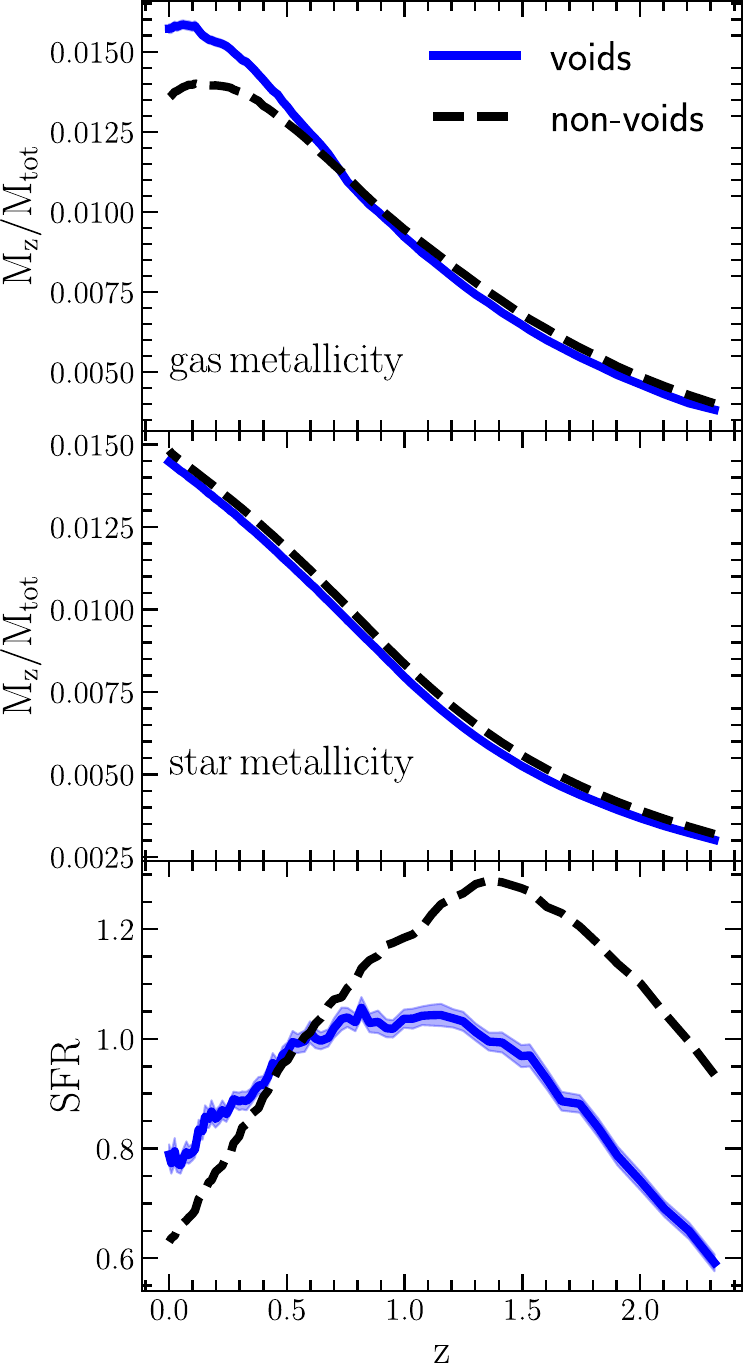}
    \caption{Evolution of different galaxy properties with redshift for galaxies in void and non-void environments, identified at $z=0$ with $10^{9.5}\, {\rm M}_{\odot} <M_{\star} < 10^{10}\, {\rm M}_{\odot}$. \textit{Upper panel:} Average gas metallicity evolution. \textit{Middle panel:} Average star metallicity evolution. \textit{Bottom panel:} Average star formation rate evolution. In all panels, the solid blue line represents the behaviour of void galaxies, while the dashed black line gives non-void galaxies. The errors, calculated using a jackknife resampling technique, are relatively small and are depicted as shaded areas for the void sample.}
    \label{fig:met_sfr_history}
\end{figure}

The evolution of the different mass components of the halos is expected to have an impact on the baryonic processes within them. Specifically, the higher potential well in void galaxies may lead to differences in the metallicity of galaxies due to a more effective retention of the gas expelled by feedback processes within the galaxies, as previously proposed by \citet{RosasGuevara2022}. To investigate this possibility, we examine the gas metallicity in Figure~\ref{fig:met_sfr_history} (top panel). The solid-blue line represents the mean relation for void galaxies, while the black-dashed line corresponds to non-void galaxies. Gas metallicity is calculated as the mass-weighted average metallicity of the gas cells bound to the subhalo, within twice the stellar half-mass radius. Metallicity is defined as the ratio of metal mass ($M_Z$) to gas mass ($M_{\rm gas}$), where $Z$ represents any element above helium (He). From the figure, it is apparent that void galaxies exhibit higher gas metallicity for $z<0.5$. In the middle panel, we present the mean stellar metallicity. The calculation methodology is analogous to that for gas metallicity, but considering stellar particles. It is noteworthy that throughout the analysed redshift range, void galaxies consistently show lower metallicities compared to non-void galaxies. Lastly, in the bottom panel, we depict the mean star formation rate. In this case, the relation reveals higher rates for void galaxies relative to non-void galaxies at $z<0.5$, while the opposite trend is observed for $z>0.5$.

We observe from the figure that void galaxies exhibit higher gas metallicity and a higher rate of star formation (at $z\lesssim 0.5$). In terms of stellar populations, void galaxies consistently show a lower metallicity across the analysed range of redshifts. The figure also illustrates that the elevated values of star formation rates (SFR) account for the disparity in the ${\rm sSFR} -M_{\star}$ relationship depicted in Figure~\ref{fig:ssfr_mst}. Since void galaxies have higher SFR values for a given $M_{\star}$ at $z=0$, their specific star formation rates (sSFR) will consequently be larger (${\rm sSFR} = {\rm SFR}/M_{\star}$).

Figures~\ref{fig:masshistory} and \ref{fig:met_sfr_history} provide further evidence that void galaxies are younger compared to the general population. This conclusion is drawn by considering the stellar mass of the galaxies as an age indicator. Furthermore, within a specific stellar mass range, we find that void galaxies have more massive dark matter haloes and exhibit a higher metallicity at $z=0$ than non-void galaxies.


\subsection{Environment comparison for an equal stellar mass distribution}
\label{sec:results_evolution2}

In the previous subsection, we showed that the evolution of the different mass components of galaxies inside of voids and outside of voids is distinct. Furthermore, this disparity has a direct impact on the astrophysical properties of galaxies, like the SFR and metallicity. Nevertheless, the intrinsic properties of galaxies are expected to be fundamentally defined by the stellar mass \citep{Dickinson2003, Baldry2006,Peng2010}. Therefore, in order to investigate whether galaxies in voids and those in other environments have differences at a fixed stellar mass, we build control samples for each redshift following this procedure: 
\begin{enumerate}
    \item We take the main branch of the merger tree of galaxies in voids having $9.5<\log(M_{\star}/{\rm M}_\odot)<10$ at $z=0$. Then, we calculate the stellar mass distribution for these objects as a function of redshift.
    \item At each redshift $z$ we randomly select galaxies outside voids in order to mimic the stellar mass distributions estimated in the previous step. These non-void galaxy samples are built in order to have 8 times more objects than the corresponding sample of void galaxies at each redshift.
    \item We calculate the mean value of different galaxy properties at each snapshot over the main branch of galaxies in voids and in the non-void samples.
    
\end{enumerate}

\begin{figure}
	\includegraphics[width=\columnwidth]{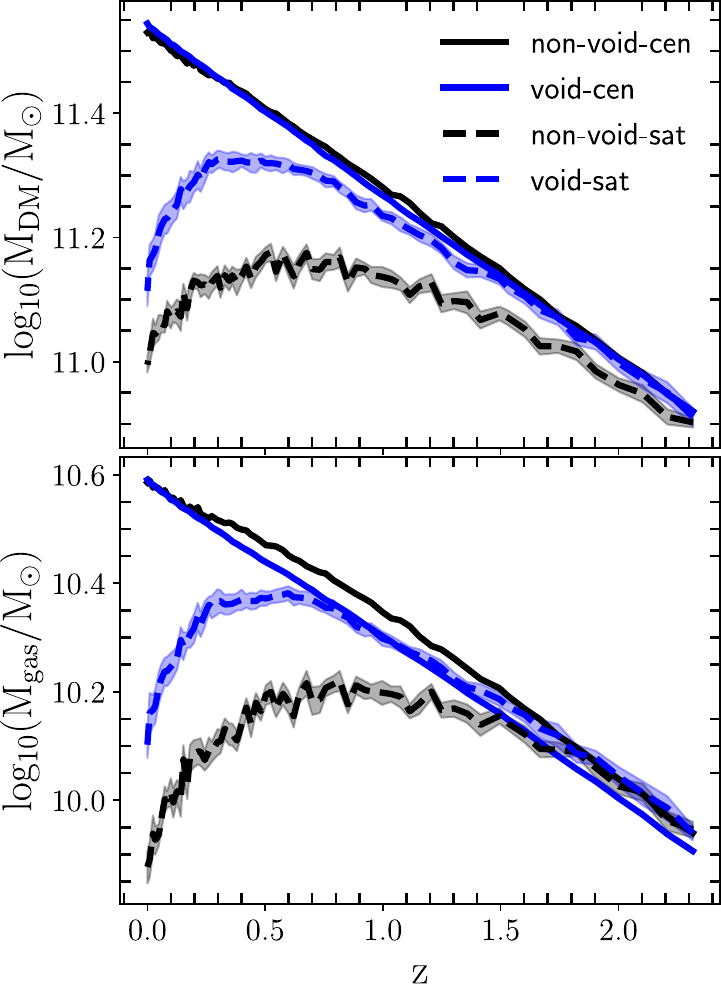}
    \caption{ In the upper panel, we show the redshift evolution of dark matter mass for central (satellite) galaxies in voids, represented by the solid (dashed) blue line. The bottom panel displays the evolution of gas mass. The black lines illustrate the mean dark matter and gas mass for galaxies in non-void environments, based on a sample with the same stellar mass distribution as the void galaxies at each redshift. The solid black line represents central galaxies, while the dashed black line depicts satellite galaxies. The galaxies in voids correspond to the same samples shown in Fig. \ref{fig:masshistory}, i.e., galaxies with $10^{9.5}\,{\rm M}_{\odot}<M_{\star}<10^{10}\,{\rm M}_{\odot}$ at $z=0$. Jackknife errors are indicated by shaded areas. 
    }
    \label{fig:masshistory_replic}
\end{figure}

Through the procedure outlined above, we ensure that the average $M_{\star}$ is the same at each time step between void and non-void galaxies, thereby mitigating potential biases arising from the observed mass differences depicted in Fig.~\ref{fig:masshistory}. The results regarding the evolution of the mean dark matter and gas masses are presented in the top and bottom panels of Figure~\ref{fig:masshistory_replic}, respectively. To assess potential variations within the local environment of galaxies, we differentiate between central and satellite galaxies. A central galaxy is defined as the subhalo with the largest number of particles/cells contained in a Friends of Friends (FOF) group. The other subhalos in the FOF groups are classified as sat;ellite galaxies. In the figures, void galaxies are represented in blue, while the general sample is depicted in black. Solid lines correspond to central galaxies, while dashed lines represent satellite galaxies. The differences in the gas and dark matter content between galaxies in void and non-void environments primarily manifest themselves in satellite galaxies, while they remain relatively small for central galaxies. We observe that, around $z\sim1$, satellite galaxies in voids begin to exhibit a lower dark matter and gas content compared to central galaxies. This suggests that this represents a characteristic redshift scale at which these galaxies are captured by other galaxies and become satellites. In contrast, non-void galaxies experience this transition at an earlier time, closer to $z\sim2$. However, it is important to note that the curves for non-void galaxies do not represent the evolution of a specific galaxy population. Instead, they reflect the mean values within a distribution designed to mimic the stellar mass distribution of galaxies in void environments.

\begin{figure}
	\includegraphics[width=\columnwidth]{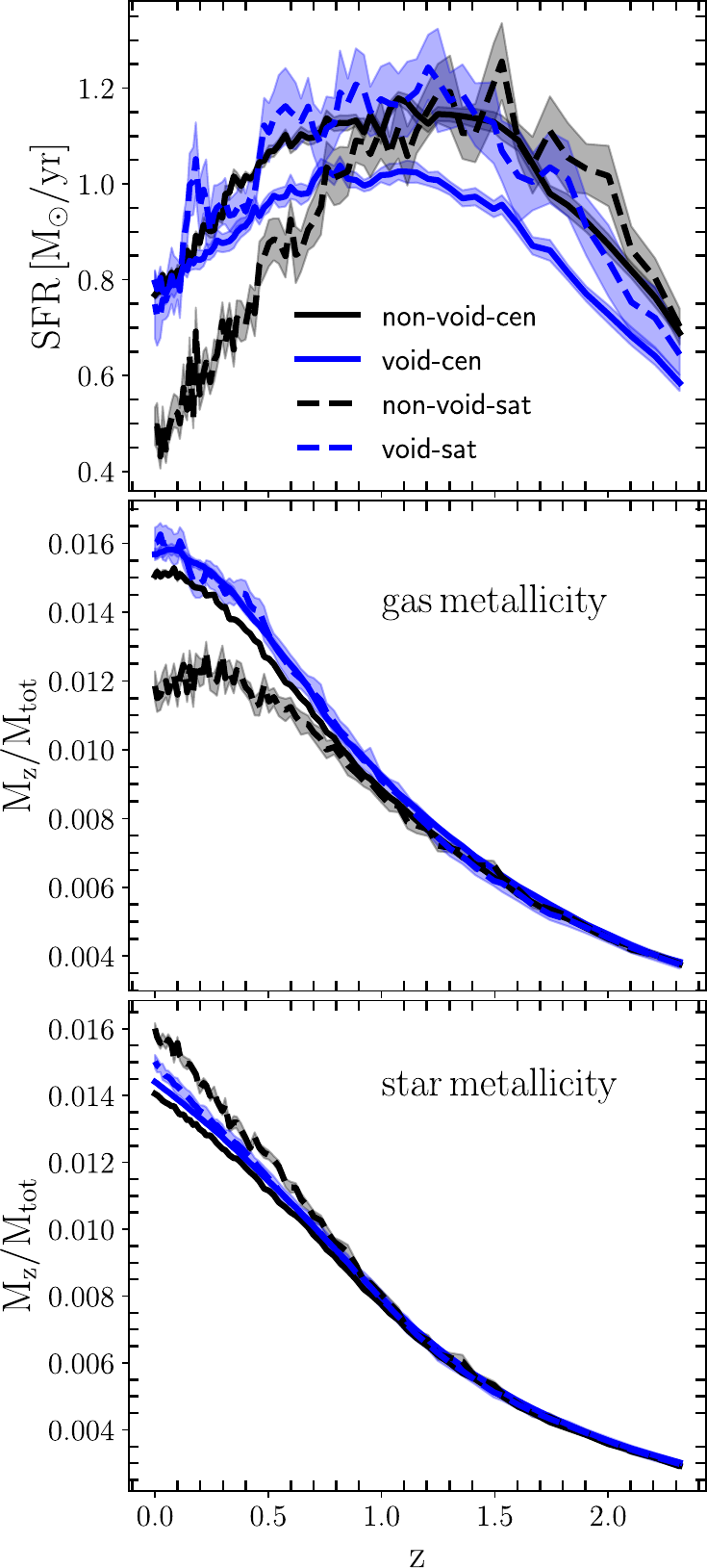}
    \caption{Similar to Fig.~\ref{fig:masshistory_replic},
    in the upper panel, we illustrate the redshift evolution of the star formation rate (SFR) for central (satellite) galaxies in voids, indicated by the solid (dashed) blue line. In the middle panel, we present the evolution of gas metallicity, and in the bottom panel, we display the evolution of stellar metallicity. The black lines give the mean SFR and metallicities for galaxies in non-void environments, based on a sample with the same stellar mass distribution as the void galaxies at each redshift. The solid black line is for central galaxies, while the dashed black line depicts satellite galaxies. Jackknife errors are shown with shaded areas.
}
    \label{fig:met-sfr0_replic}
\end{figure}

The astrophysical evolution of these galaxies is illustrated in Fig.~\ref{fig:met-sfr0_replic}. The top panel shows the star formation rate (SFR), while the middle and bottom panels depict the gas and stellar metallicity, respectively. At higher redshifts, we do not observe significant differences in the SFR between satellite galaxies in the void and non-void samples, with the SFR exhibiting an overall increase until approximately $z\sim1$. However, for $z\lesssim 1$, the SFR begins to decline, and this decrease is more pronounced in non-void galaxies. Statistically significant differences in the SFR emerge around $z\sim 1$, 
and this trend aligns with the observed gas leakage phenomenon depicted in Fig. \ref{fig:masshistory_replic}. These findings suggest a relationship between the timescale and efficiency of quenching in both sample populations. For the gas metallicity, our findings indicate that void galaxies exhibit higher levels of metallicity compared to the non-void galaxy sample at low redshifts. This disparity is more pronounced among satellite galaxies, which display significantly greater metal richness. 
On the other hand, the difference in metallicity between void and non-void central galaxies is relatively weak. However, it should be noted that void galaxies still exhibit higher metallicities. As observed in the previous figures, around $z\sim1$, the metallicities of void and non-void satellite galaxies begin to exhibit statistically significant differences, with the non-void galaxies showing lower metallicity values.

Regarding stellar metallicity, as shown in the bottom panel of Figure \ref{fig:met_sfr_history}, the differences between satellite and central galaxies in the void and non-void samples are relatively smaller. Satellite galaxies have higher metallicity than central galaxies, with non-void galaxies being more metal-rich than void galaxies. The effects on stellar metallicity begin to manifest themselves at $z\lesssim 1$. Central galaxies in both samples appear to be very similar in terms of stellar metallicity.

We would like to recall that, given the low number of particles composing galaxies at high redshift, we do not expect the galaxy properties at these redshifts to be numerically fully converged, as resolution issues may be affecting the results \citep{Pillepich2018a}. However, the qualitative trends observed do not seem to be dominated by high redshift or low stellar mass, and therefore are not dominated by resolution issues, taking also into account that we are consistently considering galaxies of the same mass in both samples.

\subsection{Mergers}
\label{sec:results_mergers}

The mechanisms by which galaxies get their stars are: \textit{in situ} star formation, that is stars are born inside the galaxy itself, and \textit{ex-situ} accretion of stars, where stars are born in other smaller galaxies that subsequently merge with the considered galaxy.  As voids are environments with a low density of galaxies, it could be expected that the rate of interactions of galaxies is smaller than in denser environments \citep{VanDeWeygaert2011}. However, numerical simulations predict that the mean number of galaxy mergers does not depend on the void environment \citep{RosasGuevara2022}. We have verified this result by comparing the mean ratio between the number of mergers in void and non-void galaxies as a function of stellar mass, without obtaining any significant environmental dependence.

\citet{RosasGuevara2022} also reported that galaxies in voids experience mergers more recently than galaxies in denser environments. To further analyse if there are differences in the time of mergers for these galaxies, we utilised the publicly available merger history catalogue of \citet{RodriguezGomez2017, Eisert2023}. This catalogue contains information regarding galaxy mergers and the \textit{ex-situ} and \textit{in-situ} stellar formation processes. 
The time of the merger is defined as the time when the corresponding merger tree branches join. The masses of the galaxies used to calculate the merger ratio are instead taken as the maximum masses reached by the progenitors prior to the merger.  
\citep{RodriguezGomez2015}
The \textit{ex-situ} stellar mass is then defined as the sum of the stellar masses of all the secondary progenitors along the whole merger history of a given galaxy.

\begin{figure}
	\includegraphics[width=\linewidth]{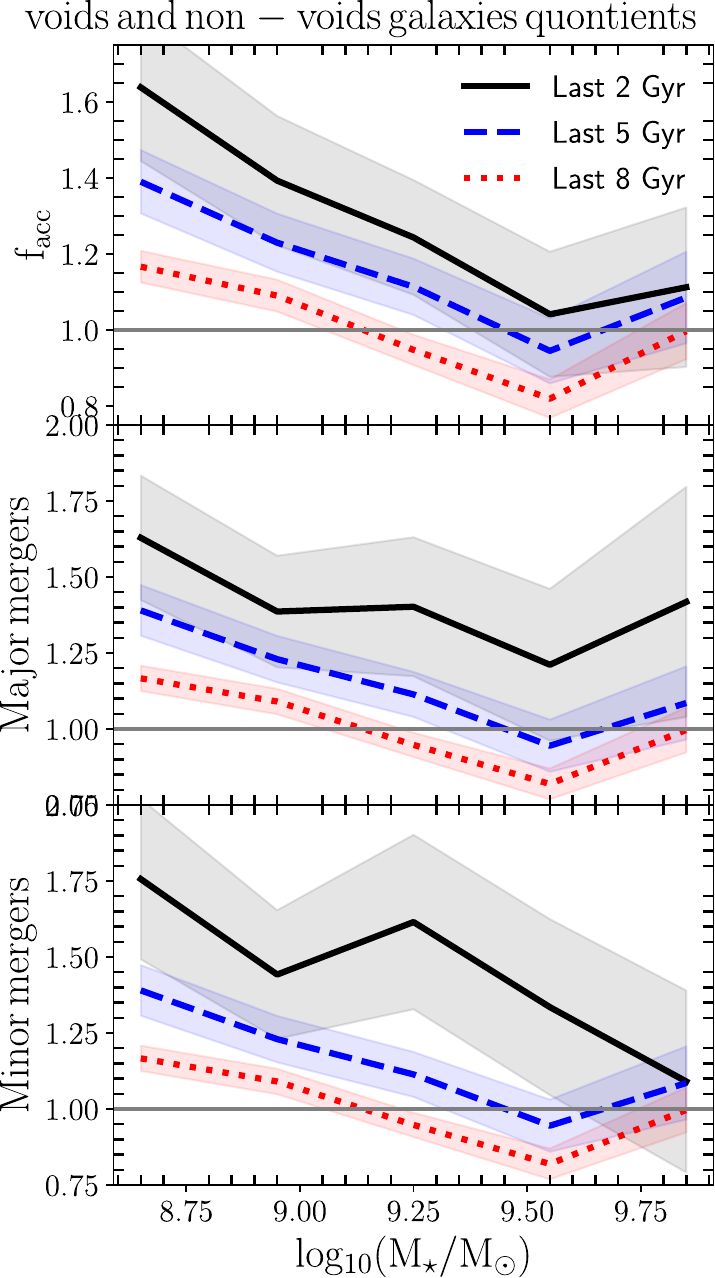}
    \caption{Mean ratio of different properties between void and non-void galaxies as a function of stellar mass. \textit{Top Panel:} the ratio of accreted mass between void and non-void galaxies. \textit{Medium Panel:} the ratio of the number of major mergers between void and non-void galaxies. \textit{Bottom Panel:} the ratio of the number of minor mergers between void and non-void galaxies. The black-solid line depicts the relationship observed over the last 2 Gyr, the blue-dashed line shows the last 5 Gyr, and the red-dotted line shows the last 8 Gyr.  We depict Jackknife errors with a shaded area.
    }
    \label{fig:mergers-acc}
\end{figure}

The top panel of Fig.~\ref{fig:mergers-acc} shows the ratio between the fraction of accreted stellar mass (\textit{ex-situ}) in void and non-void galaxies as a function of stellar mass at $z=0$. The solid line represents the ratio for the accreted mass in the last 2 Gyr ($z<0.15$), while the dashed line represents the last 5 Gyr ($z<0.5$), and the dotted line gives the last 8 Gyr ($z<1$). The dashed line illustrates the error in the ratio, estimated using the jackknife technique. We only show the error corresponding to the 2~Gyr duration since it is similar for the other two lines. The figure clearly indicates that for galaxies with $M_{\star}<10^{10.2}\,{\rm M}_{\odot}$ at $z=0$, those within voids have accreted more stellar mass recently compared to the non-void sample.

In the central panel of Fig. \ref{fig:mergers-acc}, we present the ratio between the mean number of major mergers in void galaxies and non-void galaxies as a function of stellar mass. Similarly, in the bottom panel, we show the same ratio for minor mergers. As in the upper panel, the different line styles represent 2, 5, and 8 Gyr time intervals. The figure indicates that for galaxies with $M_{\star}<10^{10.2}\,{\rm M}_{\odot}$ those within voids have experienced mergers more recently compared to non-void galaxies. This trend is observed for both major and minor mergers, with a slightly more pronounced effect for minor mergers.

Together, the panels of Fig.~\ref{fig:mergers-acc} demonstrate that void galaxies accreted mass more recently compared to non-void galaxies, at a fixed stellar mass. The additional accreted mass is a direct result of mergers. The figure clearly illustrates a prominent trend indicating that void galaxies have experienced more recent mergers. This is evidenced by an excess of approximately $50\%$ in accreted mass over the last 2 Gyr ($z<0.15$), $35\%$ over the last 5 Gyr ($z<0.5$), and $20\%$ over the last 8 Gyr ($z<1$). These values correspond to a galaxy with $M_{\star}\sim 10^{9}\,{\rm M}_{\odot}$ and are similar for galaxies with $M_{\star}<10^{10}\,{\rm M}_{\odot}$.

To investigate the differences in the total accreted mass throughout the history of galaxies, we again utilised the stellar-assembly catalogue of \citet{ Rodriguez-Gomez2016a, RodriguezGomez2016b}. This catalogue provides calculations of the accreted mass, considering all stellar particles that were not formed in the main branch of the merger tree. It is important to note that although the methodology for calculating this mass differs from that used in Fig. \ref{fig:mergers-acc}, the results obtained from both methods are consistent, as demonstrated in \citet{Rodriguez-Gomez2016a}. 

\begin{figure}
	\includegraphics[width=\linewidth]{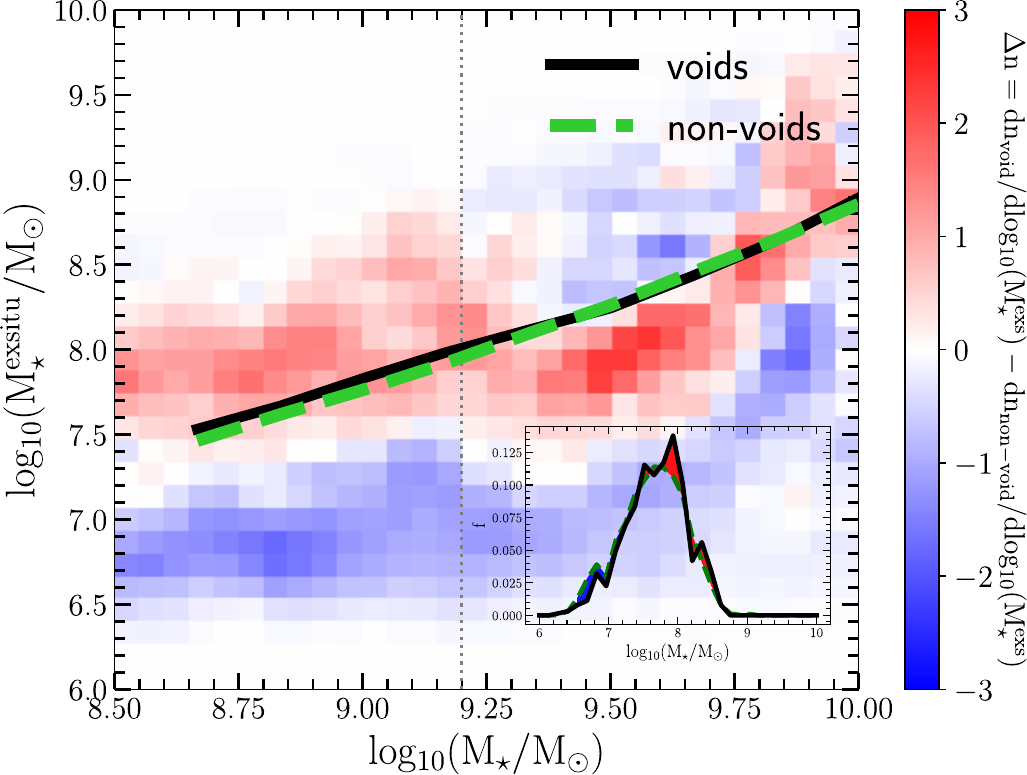}
    \caption{Mean \textit{ex-situ} stellar mass formed in galaxies belonging to void and non-void environments, plotted as a function of the total stellar mass. The continuous green line represents galaxies within voids, while the black dashed line represents galaxies classified as non-voids. The colour map shows the excess of galaxies in the \textit{ex-situ} stellar mass distribution within each total stellar mass bin. This represents the difference between the \textit{ex-situ} stellar mass distribution in void galaxies and non-void galaxies, as a function of the total stellar mass. }
    \label{fig:exsitu}
\end{figure}

In Fig.~\ref{fig:exsitu}, we present the mean accreted mass (\textit{ex-situ}) as a function of stellar mass at $z=0$. The relation for void galaxies is shown as a solid-black line, while the dashed green line represents non-void galaxies. The lines are accompanied by shaded areas that indicate the small errors in the mean, although statistically significant differences between the two lines were not found. Similar as in previous figures, a colour map is included in the plot to illustrate the difference in the distribution of accreted mass ($M_{\rm acc,\star}$) between void galaxies and the non-void sample, at a fixed stellar mass. The map is smoothed by averaging each bin with its 9 contiguous bins. 
The colour map shows that void galaxies tend to have a distribution of accreted mass skewed towards higher values.

\section{Summary and discussion}
\label{sec:discussions}

We have found consistent characteristics among the galaxies in our void galaxy sample, which align with previous studies. Our sample of void galaxies exhibits a smaller stellar-to-halo mass ratio compared to galaxies in other regions similar to previous works \citep{Tonnensen2015,Alfaro2020,Habouzit2020,RodriguezMedrano2022,RosasGuevara2022}.
Also, we report higher star formation rates \citep[as in:][]{Rojas2005,Ricciardelli2014}, and more metal-rich gas halos, while the stellar particles have lower metallicities.  
These results are consistent with recent publications from the CAVITY project, where they detected, at a given stellar mass, galaxies with lower stellar metallicities in voids compared to other environments \citep{Dominguez2023b}. 
The lower metallicities in their stars are consistent with the idea that void galaxies are younger than galaxies in denser environments \citep{Rojas2005}. However, the high gas metallicities have not been observed in previous observational studies \citep{Pustilnik2011,Kreckel2015}. It is possible that larger samples of galaxies are required to detect these signals.

Through the analysis of their assembly process, we have observed that void galaxies that end up with the same stellar mass at $z=0$ possess less stellar mass at higher redshifts, in comparison with non-void galaxies, which means that void galaxies are younger. This pattern differs when considering their gas and dark matter masses. Although they have greater gas and dark matter mass in their halos at $z=0$, these components decrease at higher redshifts. The high content of dark matter at low redshift is another manifestation of the stellar mass - halo mass relationship depicted in Fig. \ref{fig:abundance}. On the other hand, the elevated gas content of galaxies in voids aligns with previous observational studies \citep{Florez2021}.

Regarding the star formation rate (SFR) of our galaxies, we found that at higher redshifts void galaxies exhibit a smaller SFR compared to non-void galaxies. However, there is a specific period where this relationship is inverted, for instance, for galaxies with stellar masses ranging from $10^{9.5}\,{\rm M}_{\odot}$ to $10^{10}\,{\rm M}_{\odot}$ this happens at approximately $z<0.5$. This behaviour of the SFR is consistent with the star formation history (SFHs) of void galaxies constructed in \citet{Dominguez2023a}. These authors show that the SFHs are slower for galaxies within voids.
At the time of the change in the SFR behaviour, void galaxies also begin to exhibit higher metallicity in their gas. This suggests that as these galaxies reside in larger dark matter halos their gravitational potential well more effectively retains gas that is expelled with high metallicity by the feedback process of star formation.

While we observe that the differences in galaxies within different large-scale environments can be significant, these differences are primarily influenced by variations in local surroundings, as we have shown in \citet{RodriguezMedrano2023}. As illustrated in Figures~\ref{fig:masshistory_replic} and \ref{fig:met-sfr0_replic}, when studying the evolution of galaxies in the mass range of $9.5 < \log_{10}(M_{\star}/M_{\odot}) < 10$, galaxies classified as satellites exhibit variations in their dark matter and gas quantities compared to central galaxies. Central galaxies, regardless of whether they belong to voids or not, do not show substantial differences between them. On the other hand, satellites in non-void regions have the lowest gas and dark matter contents. This could be attributed to the properties of the group halos in which these galaxies reside. In voids, we do not expect to find very massive groups that could be responsible for stripping dark matter and gas from satellite galaxies. Additionally, this could also explain the low star formation rates and gas metallicities of these satellites, as in extreme cases, they may lack gas, thereby decreasing the values of these properties.
Based on these results, we can deduce that the signal in Figure \ref{fig:abundance}, which indicated that galaxies in voids prefer to reside in larger halos compared to galaxies in the non-void sample, is primarily due to satellite galaxies. As suggested, the processes of dark matter and gas stripping from galaxies with more massive structures would be more significant for the non-void sample. When we examine the dark matter content for central galaxies, we do not find differences at z=0.

As shown in Figure~\ref{fig:mergers-acc}, there is a tendency for the number of mergers in void galaxies to be higher at later times. Because merger events contribute to the assembly of stellar mass in void galaxies through the accretion process, the fraction of accreted mass in void galaxies compared with the non-void sample is higher at later times. This signal is particularly significant for very low-mass galaxies ($\sim M_{\star}<10^{9.5}$). 
Our finding that void galaxies exhibit a higher fraction of accreted mass at low redshifts is in line with the findings of \citet{Dominguez2023a}. In their study, the authors show that void galaxies assemble their stars later than those in filaments and walls, and much later than those in clusters. Given that our galaxies show high star formation rates (SFR) and experience greater accretion at low redshifts, their assembly also occurs later than in the non-void galaxy sample.
A novel result is that we find a surprisingly weak excess of accreted stellar mass in void galaxies compared to non-void galaxies, for galaxies with $M_{\star}<10^{9.25}M_{\odot}$. Although the differences in the mean accreted stellar mass are small, a consistent trend is demonstrated, as shown in Figure~\ref{fig:exsitu}. This signal challenges the common notion that void galaxies primarily grow in a merger- and interaction-free environments \citep[see for instance][]{VanDeWeygaert2011,Florez2021,Ceccarelli2022}.

We have confirmed that the local environment of where galaxies reside can be an important factor influencing their evolution. Our results indicate that galaxies classified as centrals exhibit slight differences depending on their association with voids or non-void regions, whereas more significant differences are observed for the satellite populations in both environments. This is because in denser large-scale environments (i.e., the non-void sample), massive systems of galaxies, like clusters, induce quenching processes in satellite galaxies. These systems are not present or are less massive in void regions, resulting in more moderate effects on the satellite galaxies. In this sense, the void environment modulates galactic processes, providing galaxies with a quieter environment for their evolution. Beyond these environmental effects on satellite galaxies, evolution in terms of \textit{in-situ} or \textit{ex-situ} star formation shows comparable behaviour in galaxies in both environments.

\section{Conclusions}
\label{sec:conclusions}

We identified cosmic voids in the IllustrisTNG-300 simulation and examined the formation of galaxies within these voids by analysing their merger trees. We compared the properties of these galaxies with those located outside the voids. The key findings of our study can be summarised as follows:

\begin{itemize}
\item At $z=0$, the galaxy population within voids exhibits distinct characteristics compared to galaxies outside voids with the same stellar mass. These include a higher star formation rate, a smaller stellar-to-halo-mass ratio, a higher gas metallicity, and a stellar population with lower metallicity.

\item The small observed value in the stellar-to-halo mass relation for galaxies within voids is mainly influenced by the differences between satellite galaxies in void regions and those in the control sample of non-void galaxies. When considering the evolution of satellite galaxies in voids, these objects retain more of their mass in the form of gas and dark matter compared to the satellites in the control sample.
This phenomenon can perhaps be attributed to the relative abundance of large structures, such as groups and clusters, within both the non-void and void samples.

\item The star formation rate (SFR) history of central galaxies reveals that at high redshift non-void galaxies exhibit higher rates compared to the void sample. However, the behaviour differs for the satellite population. In the non-void sample, satellite galaxies experience a sharp decline in SFR at the same redshift when they start losing gas and dark matter. Additionally, at this redshift, satellite galaxies in the non-void sample demonstrate lower gas metallicity but higher stellar metallicity compared to both central galaxies and satellite galaxies within voids.

\item When we separated the mergers that occurred within the last 2, 5 and 8 Gyr, we observed a tendency for void galaxies to experience mergers at later times compared to non-void galaxies. This surplus of mergers in void galaxies results in an excess mass accretion during these later periods.

\item The mean value in the net accreted mass differs slightly between the void and non-void samples. We observed a trend in the distribution of accreted mass, at a fixed stellar mass, towards higher values in the void sample, in the extreme of low-mass galaxies.

\item Finally, we found that the number of mergers experienced by a galaxy is not dependent on the environment. However, we did observe that galaxies within voids have undergone more recent mergers compared to galaxies in other environments, indicating a different assembly rate. Furthermore, we can conclude that although the net number of mergers appears to be the same, low-mass galaxies in voids have accreted more mass through mergers than non-void galaxies. This suggests that the nature of these mergers may be different, which deserves further investigation.

\end{itemize}

While we obtained similar results at $z=0$ compared to existing literature for galaxy properties such as the stellar-to-halo mass ratio (SHMR), star formation rate (SFR), and metallicities, we emphasise that these behaviours are predominantly influenced by the local environment, specifically the distinction between central and satellite galaxies. Therefore, we wish to underscore the importance of incorporating appropriate controls for the local environment when investigating the large-scale influences on galaxies. By accounting for the effects of the local environment, we can gain a more comprehensive understanding of the large-scale factors shaping galaxy properties.

\section*{Acknowledgements}
 
We thank the anonymous referee for a constructive and insightful report, which helped to improve the paper. The authors thank to Germán Alfaro for his valuable discussions regarding the obtained results.
This project has received funding from the European Union’s HORIZON-MSCA-2021-SE-01 Research and Innovation programme under the Marie Sklodowska-Curie grant agreement number 101086388. 
This work used computational resources from CCAD – Universidad Nacional de Córdoba (\url{https://ccad.unc.edu.ar/}), which are part of SNCAD – MinCyT, República Argentina.
This work was partially supported by the Consejo de Investigaciones Científicas y Técnicas de la República Argentina (CONICET) and the Secretaría de Ciencia y Técnica de la Universidad Nacional de Córdoba (SeCyT). ARM is doctoral fellow of CONICET. DJP and FAS of the Carrera del Investigador Científico (CONICET). The
authors thanks support by grants PIP 11220130100365CO, PICT-2016-4174, PICT-2016-1975, PICT-2021-GRF-00719 and Consolidar-2018-2020, from CONICET, FONCyT (Argentina) and SECyT-UNC.
The IllustrisTNG simulations were undertaken with compute time awarded by the Gauss Centre for Supercomputing (GCS) under GCS Large-Scale Projects GCS-ILLU and GCS-DWAR on the GCS share of the supercomputer Hazel Hen at the High Performance Computing Center Stuttgart (HLRS), as well as on the machines of the Max Planck Computing and Data Facility (MPCDF) in Garching, Germany.

\section*{Data Availability}

The void catalogue is publicy available at \url{https://catalogs.iate.conicet.unc.edu.ar/popcorn_cosmic_voids/}
The IllustrisTNG simulations, stellar assembly and merger history catalogues are publicly available and accessible at \url{www.tng-project.org/data } \citep{Nelson2019a}.  The popcorn and spherical void finders are publicly available under a MIT licence in the GitLab repository at \url{https://gitlab.com/dante.paz/popcorn_void_finder}. All other data underlying this article will be shared upon reasonable request to the corresponding authors.

\bibliographystyle{mnras}
\bibliography{example} 

\section*{Appendix: Evolution of galaxies with smaller mass}
\label{appendix}

\begin{figure}
	\includegraphics[width=\linewidth]{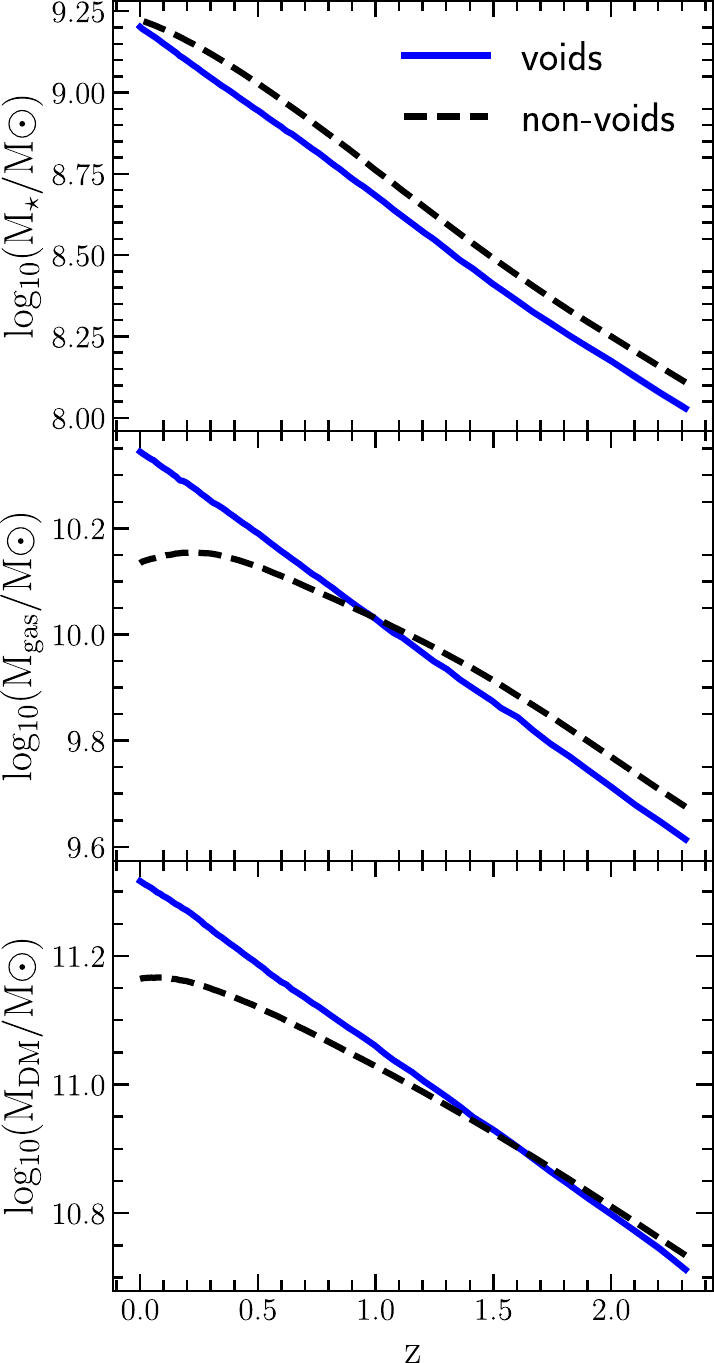}
    \caption{Evolution of different mass components with redshift for galaxies in void and non-void environments, identified at $z=0$ with $10^{9}\, {\rm M}_{\odot} <M_{\star} < 10^{9.5}\, {\rm M}_{\odot}$. \textit{Upper panel:} Average stellar mass evolution. \textit{Middle panel:} Average gas mass evolution. \textit{Bottom panel:} Average dark matter mass. In all panels, the solid blue line represents the behaviour of void galaxies, while the dashed black line is for non-void galaxies. The errors, calculated using a jackknife resampling technique, are relatively small and are depicted as shaded areas for the void sample.}
    \label{fig:masshistory_2}
\end{figure}

In this section, we present the evolution of galaxies with $9<\log_{10}(M_{\star}/M_\odot)<9.5$ at z=0. In Figure \ref{fig:masshistory_2}, the top panel illustrates the evolution of the mean stellar mass, the middle panel shows the mean gas mass, and the bottom panel displays the mean dark matter mass. The solid blue lines represent the sample of void galaxies, while the dashed black line corresponds to non-void galaxies. To account for representative uncertainty, we have included the error in the mean calculated using the jackknife technique for the void galaxy sample. As can be observed, the behavior of galaxies in this mass range is qualitatively similar to that of galaxies with $9.5<\log_{10}(M_{\star}/M_\odot)<10$, as shown in Figure \ref{fig:masshistory}.

\begin{figure}
	\includegraphics[width=\linewidth]{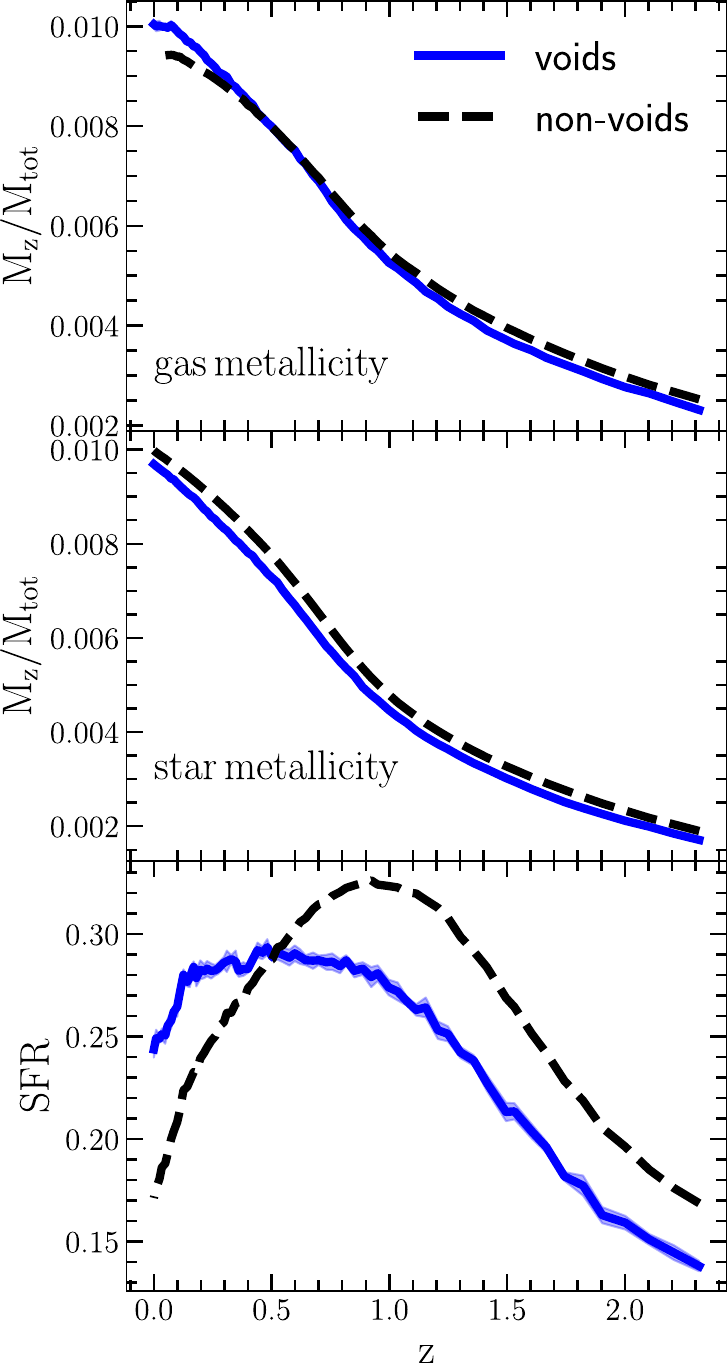}
    \caption{Evolution of different galaxy properties with redshift for galaxies in void and non-void environments, identified at $z=0$ with $10^{9}\, {\rm M}_{\odot} <M_{\star} < 10^{9.5}\, {\rm M}_{\odot}$. \textit{Upper panel:} Average gas metallicity evolution. \textit{Middle panel:} Average star metallicity evolution. \textit{Bottom panel:} Average star formation rate evolution. In all panels, the solid blue line represents the behaviour of void galaxies, while the dashed black line gives non-void galaxies. The errors, calculated using a jackknife resampling technique, are relatively small and are depicted as shaded areas for the void sample.}
    \label{fig:met_sfr_history_2}
\end{figure}

In Figure \ref{fig:met_sfr_history_2}, we present the evolution of mean gas metallicity, stellar metallicity, and star formation rate (SFR) for galaxies with $9<\log_{10}(M_{\star}/M_\odot)<9.5$ at z=0. The solid blue lines represent the sample of void galaxies, while the dashed black line corresponds to non-void galaxies. Similar to Figure \ref{fig:met_sfr_history}, void galaxies exhibit higher gas metallicity and SFR at smaller redshifts. The SFR undergoes a change in behavior near $z\sim 0.6.$ The stellar metallicity exhibits similar values in both samples.

Similar to what is indicated in sec. \ref{sec:results_evolution}, these figures show us the differences in the evolution of galaxies in voids compared to the rest of the universe. We find that, on average, galaxies in voids are younger than those in the non-void sample. These galaxies exhibit a delay in the activation of their star formation activity. Also, we show that at low redshifts these galaxies possess abundant gas and inhabit halos that are more massive than those in the rest of the universe.

\bsp	
\label{lastpage}
\end{document}